\input harvmac
\input amssym.tex

\def\title#1#2#3{\nopagenumbers\abstractfont\hsize=\hstitle
\rightline{hep-th/9805163}
\rightline{ITFA-#1}
\bigskip
\vskip 0.7in\centerline{\titlefont #2}\abstractfont\vskip .2in
\centerline{{\titlefont#3}}
\vskip 0.3in\pageno=0}
\def\authors{ \centerline{
{\bf B.J. ~Schroers}\footnote{$^{\dagger}$}
{e-mail: {\tt schroers@phys.uva.nl} }
and {\bf F.A. ~Bais}\footnote{$^{\ddagger}$}
{e-mail: {\tt bais@phys.uva.nl}}  }
\bigskip
\centerline{
Instituut  voor  Theoretische  Fysica}
\centerline{Universiteit van  Amsterdam} 
\centerline{
Valckenierstraat 65, 1018 XE Amsterdam}
\centerline{The Netherlands}
\bigskip\bigskip}
\def\abs#1{\centerline{{\bf Abstract}}\vskip 0.2in
\baselineskip 12pt
{#1}}
\def\ack{\bigbreak\bigskip\smallskip\centerline{{\bf Acknowledgements}}
\nobreak}


\font\tenbifull=cmmib10 
\font\tenbimed=cmmib10 scaled 800
\font\tenbismall=cmmib10 scaled 666
\textfont10=\tenbifull \scriptfont10=\tenbimed
\scriptscriptfont10=\tenbismall

\def\cd{\!\cdot\!}

\def\bRP{{\hbox{\bf RP}}}

\def\bk{{\hbox{\bf k}}}

\def\bI{{\hbox{\bf I}}}
\def\bJ{{\hbox{\bf J}}}
\def\bj{{\hbox{\bf j}}}

\def\bZ{{\hbox{\bf Z}}}
\def\bR{{\hbox{\bf R}}}

\def\pmatrix#1{\left(\matrix{#1}\right)}



 

\lref\HP{G. 't Hooft, ``Magnetic monopoles in unified 
gauge theories'', Nucl. Phys. {\bf B 79} (1974) 276
\semi
A.M. Polyakov, ``Particle spectrum in quantum field theory''
JETP Letters {\bf 20} (1974) 194.
}
\lref\MO{C. Montonen and D. Olive, ``Magnetic monopoles 
as gauge particles ?'', Phys. Lett. {\bf  B 72 }
(1977) 117.
}
\lref\Wittenef{E. Witten, ``Dyons of charge $e\theta/2 \pi$'',
Phys. Lett. {\bf B 86} (1979) 283.
}
\lref\GNO{ P. Goddard, J. Nuyts and D. Olive, ``Gauge theories and
magnetic charge'', Nucl. Phys. {\bf B 125} (1977) 1.
}
\lref\VaWi{C. Vafa and E. Witten, ``A strong coupling test of 
S-duality'', Nucl. Phys. {\bf B 431} (1994) 3. 
}
\lref\EW{F. Englert and P. Windey, ``Quantization condition for 
't Hooft monopoles in compact simple Lie groups'', Phys. Rev. {\bf D 14}
(1976) 2728.
}
\lref\Corrigan{E. Corrigan, ``Static non-abelian forces and the 
permutation group'', Phys. Lett. {\bf B 82 } (1979) 407.
}
\lref\SikWe{P. Sikivie and N. Weiss, ``Classical Yang-Mills theory
in the presence of external sources'', Phys. Rev. {\bf D 18} (1978) 3809
\semi
``Static sources in classical Yang-Mills theory'', Phys. Rev. {\bf D 20}
487.
}
\lref\Murray{M. Murray, ``Stratifying monopoles and rational maps'',
Commun. Math. Phys. {\bf 125} (1989) 661.
}
\lref\Godol{P. Goddard and D. Olive, 
``Charge Quantization in theories with an adjoint representation
Higgs mechanism'', Nucl. Phys. {\bf B191} 511 \semi
``The magnetic charges of stable self-dual monopoles'' Nucl. Phys.
 {\bf B 191}(1981) 528.
}
\lref\Sander{F.A. Bais, ``Charge-monopole duality in spontaneously
broken gauge theories'', Phys. Rev. {\bf D 18} (1978) 1206.
}
\lref\Abou{A. Abouelsaood, ``Chromodyons and equivariant gauge 
transformations'', Phys. Lett. {\bf B  125} (1983) 467.
}
\lref\Abouu{A. Abouelsaood, ``Are there chromodyons?'', Nucl. Phys. 
{\bf B 226} (1983) 309.
}
\lref\NC{P. Nelson and S. Coleman, ``What becomes of global color ?'',
Nucl. Phys. {\bf B 237} (1984) 1.
}
\lref\NM{P.N. Nelson and A. Manohar, ``Global colour is not
always defined'', Phys. Rev. Lett. {\bf 50} (1983) 943.
} 
\lref\BMMNSZ{
A. Balachandran {\it et al.}, ``Monopole topology and the problem of color'',
 Phys. Rev. Lett {\bf 50} (1983) 1553.
}
\lref\Swansea{N. Dorey, C. Fraser, T.J. Hollowood and M.A.C. Kneip,
``Non-abelian duality in N=4 supersymmetric gauge theories'', 
{\tt hep-th/9512116}
 \semi
``S-duality in N=4 supersymmetric gauge theories with arbitrary
gauge group'',Phys. Lett.{\bf  B 383} (1996) 422.
}
\lref\sen{A. Sen, ``Dyon-monopole bound states, self-dual harmonic
forms on the multi-monopole moduli space, and $SL(2,\bZ)$ invariance 
in string theory'', Phys. Lett. {\bf B 329 } (1994) 217. 
}
\lref\abduality{G. Segal and A. Selby, ``Cohomology of spaces of monopoles'',
Commun. Math. Phys. {\bf 177} (1996) 775 \semi
J. Gauntlett and D. Lowe,
``Dyons and S-duality in N=4 supersymmetric gauge theory'',
 Nucl. Phys.{\bf  B472} (1996) 194 
\semi
 K. Lee, E.J.  Weinberg and  P. Yi, ``Electromagnetic duality and SU(3)  
monopoles'' Phys. Lett.{\bf B  376 } (1996) 97
\semi
G.W. Gibbons, ``The Sen conjecture for fundamental monopoles of distinct
type'', Phys. Lett. {\bf B 382 } (1996) 53. 
}
\lref\indrep{A.O. Barut and R. Raczka, {\it Theory of group representations
and applications}, PWN, Warszawa 1977.
}
\lref\cgcoeff{J.S. Rno, ``Clebsch-Gordan coefficients and special functions
related to the Euclidean group in three-space'', J. Math. Phys. {\bf 15}
(1974) 2042
\semi
N.P. Landsman, ``Non-shell unstable particles in thermal field theory'',
Ann. Phys. {\bf 186} (1988) 141.
}
\lref\threerus{D.A. Varshalovich, A.N. Moskalev and V.K. Khersonskii, 
{\it Quantum Theory of Angular Momentum}, World Scientific, Singapore 1988.
}
\lref\BDP{F.A. Bais, P. van Driel and M. de Wild Propitius, ``Quantum
 symmetries in discrete gauge theories'', Phys. Lett. {\bf B 280} (1992) 63
\semi
 M. de Wild Propitius and  F.A. Bais, ``Discrete gauge theories'',
Proc. CRM-CAP Summer School ``Particles and fields '94'', Banff, Springer
Verlag (1996),  {\tt  hep-th/9511201}.
}

\lref\HR{P.A. Horv\'athy and J.R. Rawnsley, ``Internal symmetries of 
non-abelian gauge field configurations'', Phys.Rev. {\bf D 32} (1985) 968
\semi
``The problem of ``global colour'' in gauge theories'', J.Math.Phys {\bf 27}
(1986) 982
}
\lref\BS{F.A. Bais and B.J. Schroers, ``Quantisation of monopoles with
non-abelian magnetic  charge'' Nucl.Phys. {\bf B 512} (1998) 250
}
\lref\KS{J.R. Klauder and B.-S. Skagerstam, {\it Coherent states}, 
World Scientific, Singapore 1985.
}
\lref\Perelomov{A.M. Perelomov, ``Coherent states for 
arbitrary Lie groups'', Commun.Math.Phys. {\bf 26} (1972) 222.
}

\title{98-12}
{S-duality in SU(3) Yang-Mills theory }
{ with non-abelian unbroken gauge group}

\authors
 
\abs
\noindent 
It is observed that the magnetic  charges of classical monopole solutions
in Yang-Mills-Higgs  theory with non-abelian  unbroken gauge group $H$ 
are in  one-to-one correspondence with coherent states of a  dual or 
magnetic  group $\tilde H$. In the spirit of the Goddard-Nuyts-Olive
conjecture  this observation  is interpreted as evidence  for a hidden 
magnetic symmetry of Yang-Mills theory. $SU(3)$ Yang-Mills-Higgs theory with
unbroken gauge group $U(2)$ is studied in detail. The action of the magnetic 
group on semi-classical states is given  explicitly. Investigations of 
dyonic excitations  show that  electric and magnetic symmetry  are never 
manifest at the same time: Non-abelian magnetic charge obstructs the  
realisation of electric symmetry and vice-versa. On the basis of this fact  
the charge sectors in the theory  are classified and their fusion rules
are discussed.  Non-abelian electric-magnetic  duality is formulated as a map 
between charge sectors. Coherent states  obey particularly simple fusion 
rules, and in the set  of coherent states $S$-duality can  be formulated 
as an $SL(2,\bZ)$ mapping  between  sectors which leaves the fusion rules 
invariant. 

\hbox{}

\centerline{\it to appear in Nuclear Physics B}

\Date{May 1998}

\baselineskip 16pt
\newskip\normalparskip
\normalparskip = 4pt plus 1.2pt minus 0.6pt
\parskip = \normalparskip

\newsec{Outline of the problem}

Yang-Mills-Higgs theory with gauge group $G$ spontaneously broken 
to a subgroup $H$  contains two sorts of particles. 
The perturbative particles in the quantised
theory  can  be organised into unitary irreducible representations  (UIR's)
of $H$ and, in terminology borrowed from the abelian case $H=U(1)$,
 the particles in non-trivial UIR's
may be called ``electrically charged''.
It is well-known that spontaneously broken Yang-Mills-Higgs theory also
contains magnetically charged particles if $\pi_2(G/H)$ is non-trivial. 
These arise as  
solitonic solutions of the classical Euler-Lagrange  equations.
As a consequence of the  generalised Dirac  quantisation condition
the magnetic charges are quantised and take values in  a certain lattice
\EW. It was emphasised by Goddard,
Nuyts and Olive (GNO)  \GNO\ that one  could interpret this lattice as the 
weight lattice of a dual or magnetic group $\tilde H$.
GNO conjectured that   the presence of the magnetic monopoles
signals a  hidden magnetic symmetry of Yang-Mills theory, and that 
the full symmetry is the product $\tilde H\times   H$.
This  conjecture was further elaborated  by Montonen and Olive \MO\ in the 
case where $H=U(1)$. According to the Montonen-Olive electric-magnetic
duality  conjecture
 the physics of the electrically charged particles 
at coupling $e$ is the same as that of the magnetically charged particles 
at coupling $4\pi/e$.  More recently  the  generalisation of this conjecture
to so-called  $S$-duality by  Sen \sen\ has attracted much attention. 
The picture that emerges from Sen's work and the evidence to support 
his conjecture \abduality\  is that ${\cal N}=4$ supersymmetric
 Yang-Mills  theory
with abelian unbroken gauge group $H$  contains
electric, dyonic and magnetic charge sectors, and that these are mapped into
each other by the duality group $SL(2,\bZ)$.
 
In this paper we continue our investigation of charged excitations
in Yang-Mills-Higgs  theory with non-abelian
unbroken gauge symmetry, begun in \BS.
Our goal is to   understand  the structure of 
the charge sectors   and to formulate $S$-duality in this 
setting. Precisely we 
consider Yang-Mills theory on (3+1)-dimensional Minkowski space,
with gauge group $G$ and complex coupling 
$\tau = {\theta \over 2 \pi } +  {4\pi i\over  e^2}$
(combining the  coupling constant $e$ with the $\theta$-angle 
of the theory). We will 
mostly consider the ${\cal N}=4$ supersymmetric version of 
the theory here. Although we will not  perform any explicitly supersymmetric
computations, embedding our arguments in the ${\cal N}=4$ supersymmetric
setting allows us to make certain quantitative statements.
Since the ${\cal N}=4$ theory has a vanishing $\beta$-function
we are in particular able to refer to a scale independent
 coupling  constant.  We should stress,
however,  that we expect the qualitative aspects of the picture we 
are going to present to remain valid even in the non-supersymmetric situation.

To formulate the problems we have to address as clearly as possible,
we briefly review the abelian situation. 
All the relevant 
features are present in the simplest model, that of $G=SU(2)$ broken
to $H=U(1)$. In that theory the electric charge of an excitation is
given by a single integer $N$, the label of an UIR of $U(1)$. The 
magnetic charge also takes integer values, which we denote by  $K$.
While the mathematical status of that  integer is topological (it
is the degree of the Higgs field at spatial infinity) it can fruitfully
be interpreted as a representation label of a magnetic $U(1)$.
To distinguish the electric from the (hypothetical) magnetic  $U(1)$
we  write  $\tilde U(1)$
for the latter.   A general
dyonic sector  may  thus  be characterised by a  pair  of integers $(K,N)$.
The important point here is that the fusion rule for dyonic sectors
\eqn\abfus{
(K_1,N_1)\otimes(K_2,N_2) = (K_1+K_2, N_1+N_2)
}
is indeed the Clebsch-Gordan series for representations 
of  $\tilde U(1)\times U(1)$.
While this remark is a triviality in the abelian case, we will see that 
the fusion properties of dyonic sectors impose severe constraints 
on the group-theoretical interpretation of magnetic charges in the non-abelian
context. 

The duality  group in this case is  $SL(2,\bZ)$, which acts naturally
 on the  pair $(K,N)$. Thus the element  
\eqn\sltwoz{
\left(\matrix{a & b\cr c & d}\right)
\in SL(2,\bZ)
}
maps the sector $(K,N)$ onto the sector $(K,N)M^{-1}=(dK-cN,-bK+aN)$ while
transforming simultaneously the coupling $\tau$ via a modular transformation
to 
$(a \tau  + b) /( c\tau +d )$. In particular the element
\eqn\modua{
S = \left(\matrix{0 & 1\cr -1 & 0}\right)
}
implements the Montonen-Olive electric-magnetic (and weak-strong  coupling)
duality \MO, and the element
\eqn\witeff{
T=\left(\matrix{1 & -1 \cr 0 & 1}\right)
}
implements the Witten effect (the $2\pi$-shift in the $\theta$-angle) 
\Wittenef.

Finally we emphasise 
that the $SL(2,\bZ)$ action on 
the sectors respects  the fusion rule \abfus.  In fact one can
invert the logic and  ask which permutation of the sectors $(K,N)$ is an 
automorphism of the fusion rules. Any permutation $\Pi$  of the sectors can
be  expressed in terms of an  invertible map  $\pi: \bZ^2\rightarrow
\bZ^2$ of the integer labels $(K,N)$. The requirement 
\eqn\fusinv{
\Pi((K_1,N_1))\otimes \Pi((K_2,N_2)) = \Pi((K_1+K_2, N_1+N_2))
}
means that the associated map $\pi$ is linear. Since  invertible
linear maps $\bZ^2\rightarrow \bZ^2$  constitute the group $SL(2,\bZ)$
one  could  define the $S$-duality group also as the automorphism group
of the fusion ring. This point of view will play an important role
in our discussion.

There are  two principal problems
which one encounters 
when trying to generalise the above story to the non-abelian regime.
The first concerns  the identification of the magnetic group. 
While the interpretation of the magnetic charge lattice as the 
weight lattice  of the magnetic group goes back to \GNO, the 
precise identification of monopole solutions one finds classically
with the UIR's of the conjectured magnetic group has so far not 
been achieved, despite various efforts \Swansea.

The second problem concerns the dyonic sectors of the theory.
It was  first noticed by Nelson and Manohar \NM, further elaborated
by Horvathy and Rawnsley \HR\ and more recently by us \BS\ that in
the presence of non-abelian magnetic  charge   only that part of the unbroken 
(electric) group  has a globally defined action 
 on  a classical configuration which commutes with 
the magnetic charge.
Thus, unlike in the abelian case, we cannot expect to label dyonic
sectors by UIR's of the product of the electric and the magnetic group.
Rather we  require a labelling which 
 accounts for the  interplay between magnetic and electric
charges.

We offer solutions  to both 
these problems  here. The essential input which allows us to overcome
the first of the above problems is the interpretation of the classical
monopole solutions as coherent states of the magnetic group.
Putting this together with the results of our earlier paper \BS\
 we present a consistent labelling of the magnetic, dyonic and 
electric sectors of the theory. 
Purely electric sectors are labelled by UIR's of the electric group
$H$ and purely magnetic sectors by UIR's of the magnetic group $\tilde H$,
but  the  important point is that electric and magnetic symmetry are
never simultaneously manifest: one can at most implement 
subgroups of $H$ and $\tilde H$ which, in a sense to be specified in 
this paper, commute with each other.

The fusion rules  of the sectors are 
considerably more  intricate than in the abelian case and  
depend on the coupling regime; they were discussed in \BS\
for the case of  weak electric coupling $e\ll 1$.
Here we are able to use our evidence for the dual group $\tilde H$
to understand the fusion rules also  in the strong electric 
coupling regime $e\gg 1$. Our picture includes a truly non-abelian
implementation of electric-magnetic duality. Moreover we find that, if we 
restrict attention to coherent states (both magnetically and electrically),
the fusion rules simplify and $S$-duality can again be formulated as 
an $SL(2,\bZ)$-action on   charge sectors which leaves the fusion rules
invariant.

In this paper we shall explain our ideas in Yang-Mills theory with 
gauge group $SU(3)$ broken to $U(2)$. This is the simplest model 
which displays all the phenomena we want to discuss. 
Our main reference
throughout is the paper \BS, where the semi-classical properties of 
monopoles in this model were discussed in detail. Nonetheless
the present paper can be read independently; whenever 
results from \BS\ are used we have stated them carefully.

\newsec{Classical monopoles as coherent states of the magnetic group}

\subsec{Classical Monopoles}

The bosonic fields of ${\cal N}=4$ supersymmetric Yang-Mills  theory with 
coupling constant $e$ are a connection and an adjoint  Higgs field. 
A classical (static) monopole solution is a pair $(A_i,\Phi)$
of such a connection  (we work in the temporal gauge $A_0=0$) and a Higgs
field on $\bR^3$ satisfying the Bogomol'nyi equation:
\eqn\bog{\eqalign{
D_i\Phi = B_i
}} 
as well as certain boundary conditions.
Here 
$D_i= \partial_i + e A_i$ is  the covariant derivative
and $B_i$ is  the non-abelian magnetic field
\eqn\magn{\eqalign{
B_i = {1\over 2} \epsilon_{ijk}\left(\partial_j A_k -\partial_k A_j
+ e [A_j, A_k]\right).
}}
For  details about the boundary conditions and the 
 notational conventions concerning the gauge group $SU(3)$
we refer the reader to \BS. Here we note only those boundary
conditions which concern the symmetry breaking and the magnetic  charge.
The symmetry breaking scale is set by 
\eqn\syca{
-\tr \,\Phi^2 \rightarrow {1\over 2} v^2 \quad \hbox{for} \,\, r \rightarrow
\infty.
}
Further  we demand that 
the  Higgs field has the following form 
along the positive z-axis:
\eqn\hasy{\eqalign{
\Phi(0,0,z) = \Phi_0 - {G_0 \over 4 \pi  z} +{\cal  O}({1\over z^2}),
}}
where   $\Phi_0$ and  $G_0$ are   constant elements of the Lie algebra 
$su(3)$.  The former  determines the symmetry breaking pattern
and may be chosen to lie
in the Cartan subalgebra.
For the minimal symmetry breaking case we are interested in we require
$\Phi_0$ to have one repeated eigenvalue, so we take
\eqn\symbr{
\Phi_0 = {iv\over 2 \sqrt 3} \pmatrix{1  & 0 & 0 \cr 0 & 1 & 0 \cr 0 & 0 & -2}.
}
Then the generators of the unbroken $U(2)$ symmetry
have the following form at $z=+\infty$:
\eqn\gens{\eqalign{
I_1& = {1\over 2} \pmatrix{0 &1  &0\cr1  & 0  & 0 \cr0 & 0 & 0}, \,\, 
I_2=  {1\over 2} \pmatrix{0 &-i &0\cr i  & 0  & 0 \cr0 & 0 & 0}, \,\,
I_3=  {1\over 2}\pmatrix{1  & 0 & 0 \cr 0 & -1 & 0 \cr 0 & 0 & 0} \cr
Y& =  {1\over 3} \pmatrix{1  & 0 & 0 \cr 0 & 1 & 0 \cr 0 & 0 & -2}
}}

The constant Lie algebra element 
$G_0$ is the magnetic charge of the  configuration and 
has to satisfy the generalised  Dirac quantisation condition \EW, \GNO:
\eqn\dirac{
\exp{(e G_0)} = 1.
}
The magnetic charge $G_0$  may also be rotated into the Cartan subalgebra,
and then the Dirac condition forces it to lie on a certain lattice, 
the dual root lattice of $su(3)$. However, in the case of minimal
symmetry breaking it is not natural to require $G_0$ to lie in Cartan
subalgebra.
A better way to characterise the magnetic charge is to consider the  orbit
of $G_0$  under the action of the  gauge group $U(2)$, acting 
in the base point $(0,0,\infty)$. As shown in \BS\ these orbits
are either trivial, in which case they are points of quantised ``height''
in the Lie algebra $su(3)$,  or  two-spheres  
of quantised radius and ``height'' in $su(3)$.  These orbits are thus
characterised by two numbers  $K$ and $k$, where $K$ (the ``height'') 
is an integer  and $k$ (the radius)  is a non-negative  half-odd integer 
if $K$ is odd and 
a non-negative  integer if $K$ is even; see Fig.~1. 
Explicitly, each
non-trivial  orbit 
can be parametrised by spherical coordinates $(\alpha,\beta)\in [0,2\pi)\times
[0,\pi)$ so that 
an element of the  orbit labelled by $(K,k)$ can be written as 
\eqn\magorb{
G_0(\alpha,\beta) = {4 K \pi i  \over e }
\left( { 3\over 4} Y \right) + 
{4\pi i \over e} \bk\cd \bI,
}
where the vector $\bk=(k_1,k_2,k_3)$ has   length 
$|\bk| = k$ and the  direction $\hat \bk$  parametrised  
by $(\alpha,\beta )$:
\eqn\magvec{
\hat \bk = (\sin \beta \cos \alpha, 
\sin \beta \sin \alpha, \cos \beta).
}
We emphasise that the coordinates $(\alpha,\beta)$ stem from the action
of the unbroken  gauge group $U(2)$ at one point only. As mentioned 
in  Sect.~1,  this action cannot be extended to a global action in the
presence of non-abelian magnetic charge. Clarifying  the physical 
interpretation of the magnetic orbits and their coordinates is 
one of the goals of this section.

To sum up, the allowed magnetic charges of a monopole can  be written as a pair
$(K,\bk)$ consisting of an integer $K$  and a vector
$\bk $ of quantised length
$k$ such that $K+2k\in 2 \bZ$. As explained in  \BS\ it follows from
 the results of \Murray\ that 
for  solutions of the Bogomol'nyi equations the charges
necessarily lie inside the cone $k\leq |K|/2$.
 
\vskip .6in
\input epsf 
\epsfxsize=10truecm
\centerline{
\epsfbox{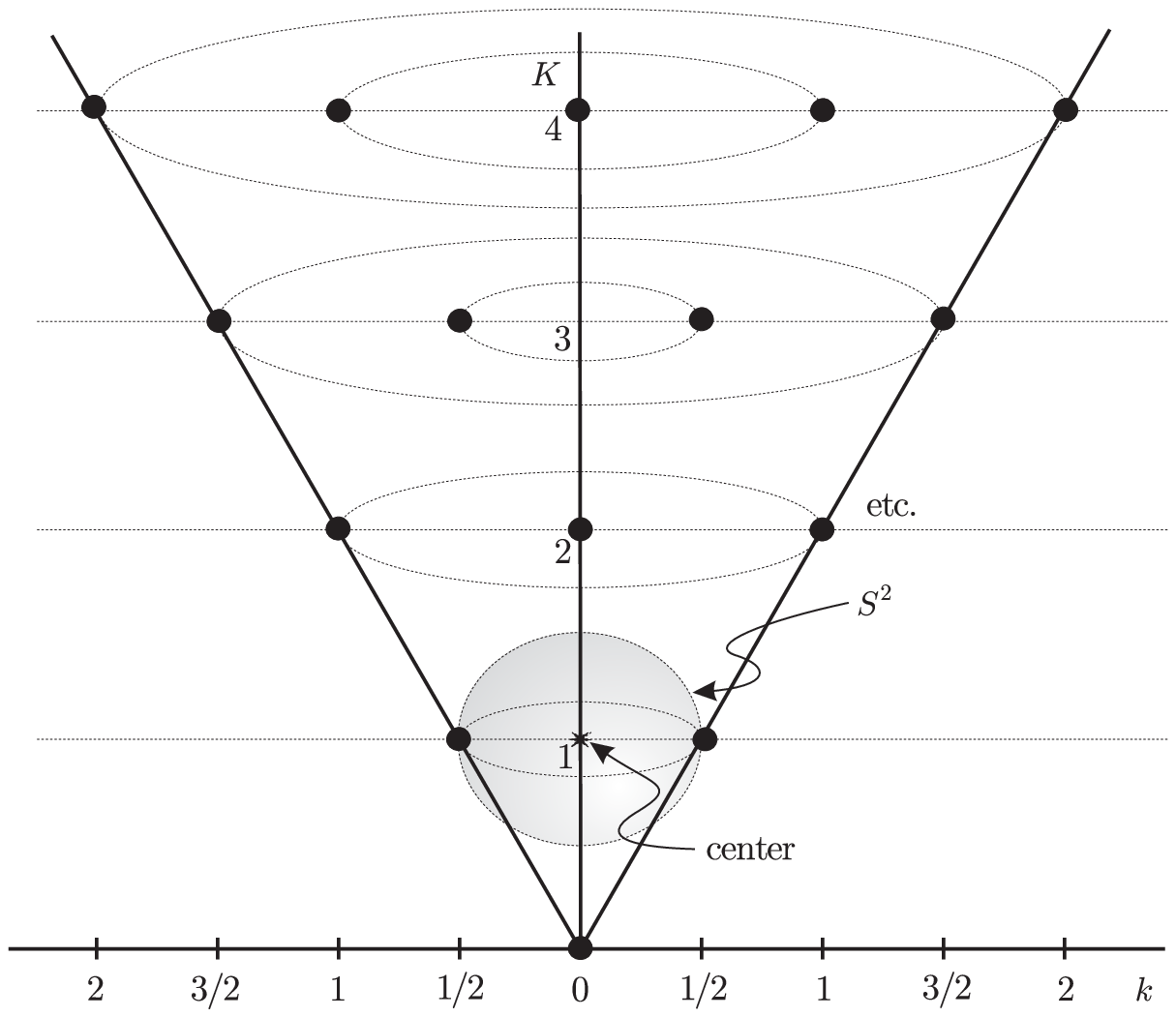}
}
\bigskip

{\footskip14pt plus 1pt minus 1pt \footnotefont
\noindent {\bf Fig.~1:} {\it   The magnetic orbits classifying 
 $SU(3)$ monopoles with minimal symmetry breaking}
}
\bigskip

We are now able to phrase in more precise terms 
 the  first of the two principal problems
outlined in the introduction,  that of  identifying the magnetic group.
By definition that group should have UIR's which classify the 
magnetic monopoles found in the theory.
The challenge is thus to identify the  classical set of monopoles whose charges
lie on an orbit  with labels  $K$ and $k$ with the UIR of some group.  
We shall now show how  this can be achieved and how a magnetic group
 can be defined. The key step is the interpretation of  the 
classical monopole charges as labels of coherent states.
  In anticipation of this result we  
introduce the notation
\eqn\mopost{
|K;k,\hat \bk\rangle
}
for monopole charges, where we have separately notated the length
and the direction of the  non-abelian charge $\bk$.

\subsec{Coherent states revisited}

The best-known coherent states  are related to  the Heisenberg group,
but various generalisations  have been studied in the literature.
For us  the concept of  coherent states for general  
compact Lie groups introduced in \Perelomov\ is particularly relevant. 
In this paper we mainly need 
coherent states of $SU(2)$ which are also much discussed in 
the literature,  usually  under the names ``Bloch states'' or 
``spin coherent states''.  For a guide to the vast literature on the
subject we refer the reader to the collection of reprints \KS.
There exist various different conventions for defining
coherent states, mostly to do with choosing a ``fiducial state''.
We consider coherent states defined in terms of a highest weight state,
but in order to avoid confusion we  will carefully state  our conventions.
Thus  we introduce
 Euler angles $(\alpha,\beta,\tilde \gamma) \in [0,2\pi) \times [0,\pi)
\times [0,4\pi)$ by writing an  $SU(2)$ matrix $P$ in terms of the
Pauli matrices $\tau_1,\tau_2$ and $\tau_3$ as 
\eqn\euler{
P(\alpha,\beta,\tilde \gamma) =  e^{-{i\over 2} \alpha \tau_3}
e^{-{i\over 2} \beta \tau_2}
e^{-{i\over 2} \tilde \gamma \tau_3}.
}
The coherent state $|k, \hat \bk\rangle$  
in  a spin $k$ representation $V_k$   of $SU(2)$ with basis
$\{ |k,m\rangle \}$ ($m=-k,-k+1, ...,k-1,k$) is  defined in terms  
of the highest
weight state $|k,k\rangle$ as 
\eqn\coherent{
|k,\hat \bk \rangle = D^k(P)|k,k\rangle = \sum_{m=-k}^k D^k_{mk}(P) 
|k,m\rangle,
}
where $D^k(P)$  is the representation of $P$ in $V_k$
and 
\eqn\wigner{
D^k_{ms}(P)  = 
\langle k,m|D^k(P)|k,s\rangle  = e^{-im\alpha} d_{ms}^k (\beta) 
e^{-is\tilde \gamma}
} 
are the Wigner functions in the conventions of \threerus. 
The coherent  states are labelled by the total  spin $k$ and 
the  unit vector $\hat \bk$, which  is the image
of $P\in SU(2) \sim S^3$ under the Hopf projection
\eqn\Hopf{
\hat \bk = \pi_{\footnotefont{Hopf}}(P),
}
where $\hat \bk$ is given in terms of the Euler angles $(\alpha,\beta)$
as in    \magvec. Coherent states  
 only depend on the angle $\tilde \gamma$ via a phase factor
and it is conventional in the discussion of coherent states to choose
$\tilde \gamma =0$.

The expectation value of the spin operator $\bJ = (J_1,J_2,J_3)$
has the following simple  form for coherent states: 
\eqn\spinex{
\langle k,\hat \bk|\bJ|k ,\hat \bk\rangle = k\hat\bk.
}
In particular the length of the expectation value of the 
spin operator is thus equal to the total spin $k$  for a coherent state.
This property can also be used as the defining property of 
coherent states. This approach gives a better conceptual understanding
and we want to adopt it here although it does not seem to be 
standard in the literature. To appreciate the significance of the 
definition we are going to give it is useful to recall some simple
facts about $SU(2)$ representations.

\noindent {\it Fact 1:}  Let $|\psi\rangle$ be a vector in the carrier 
space $V_k$  of the spin $k$ representation of $SU(2)$. Then the expectation
value $\bj_{\psi} = \langle \psi | \bJ |\psi\rangle$ of the spin operator has
length at most $k$:   $|\bj_{\psi}| \leq k$.

To see this use the relations 
\eqn\standi{\eqalign{
J^{\hat k}_+J^{\hat k}_- &= \bJ^2 - (\hat \bk\cd\bJ)^2 +\hat\bk\cd \bJ\cr
J^{\hat k}_-J^{\hat k}_+ &= \bJ^2 - (\hat \bk\cd\bJ)^2 -\hat\bk\cd \bJ
}}
for an arbitrary  quantisation axis  $\hat \bk$ and 
operators $J^{\hat k}_+$ and $J^{\hat k}_-$ which act as raising and 
lowering operators in the spectrum of $\hat \bk\cd \bJ$. Computing the 
expectation value of both sides for an arbitrary state $|\psi\rangle \in V_k$
 and using the positivity of the 
resulting LHS one deduces
\eqn\funt{
k(k+1) \geq |\hat\bk\cd\bj_\psi|(|\hat\bk\cd\bj_\psi| + 1)
\Rightarrow k \geq  |\hat\bk \cd \bj_\psi| .
}
Since this holds for all $\hat \bk \in S^2$ the claim follows.

An immediate corollary is the

\noindent {\it Fact 2:} For all $|\psi\rangle \in V_k$ 
the  square of the ``uncertainty''  of the spin vector
$\sigma_{\psi}^2 (\bJ) = \sum_{a=1}^3 \langle \psi| J_a^2|\psi\rangle
- \langle \psi |J_a |\psi\rangle ^2 $ is bounded below by $k$.

It is easy to check that the bounds on the length of $\bj_{\psi}$ and the 
variance $\sigma_{\psi} (\bJ)$ are attained for  the coherent states
\coherent. Conversely we can define coherent states of $SU(2)$ to
be those states   $|\psi\rangle$
in  the carrier space $V_k$ of the spin $k$ representation  for which
the length of expected spin vector $\bj_{\psi}$  is  maximal ($=k$)
and the variance $\sigma_{\psi} (\bJ)$ is minimal ($=\sqrt k$).
A coherent state $|\psi\rangle \in V_k$
 can thus be characterised uniquely    by the direction $\hat \bk$ 
of $\bj_{\psi}$. In this approach, the equation   \spinex\ 
is  the defining equation for the coherent state
  $|k,\hat\bk\rangle$.

Coherent states are over-complete. In 
particular  the  inner product $  \langle k, \hat \bk'| k, \hat \bk\rangle$
of two coherent states  
 in the spin $k$ representation only  vanishes between the coherent states
associated with antipodal points, i.e. if   $\hat \bk' =  - \hat \bk$. 
Finally we note   the following decomposition of the identity
in terms of coherent states in the carrier space $V_k$:
\eqn\ide{
\hbox{Id} ={2k+1 \over  4 \pi} 
\int \sin \beta d\beta d\alpha \,\,|k,\hat \bk\rangle \langle k,\hat\bk|.
}

\subsec{The magnetic group}

To make contact between  magnetic charges and  coherent states we
need to include the $U(1)$-factor  of $U(2)=
\left( U(1) \times SU(2) \right ) / \bZ_2$ in our picture. All 
$U(1)$ representations are one-dimensional and a normalised basis state
$|K\rangle$  in
the  charge $K$ representation of $U(1)$ is automatically coherent. 
We denote  the carrier space of an  UIR  of $U(2)$ by $V_{K,k}$, where
$K$ is the $U(1)$  charge and $k$ the $SU(2)$ spin and we have the 
the constraint $K+2k \in 2 \bZ$ in order to respect the $\bZ_2$
identification. Then a   coherent state in  $V_{K,k}$  is of the form
\eqn\utwoco{
|K;k,\hat\bk\rangle = |K\rangle \otimes |k,\hat\bk\rangle.
}
It  is  clear from our notation that there is a 
natural bijection between  monopole charges \mopost\ 
 and coherent states \utwoco\  of $U(2)$. 
Here we propose to identify the two. 
Thus we interpret the charges $(K,k)$ characterising the magnetic 
orbits as  labels of UIR's  of a magnetic copy of $U(2)$.
To distinguish this $U(2)$  from the unbroken electric group $U(2)$
we denote it by   $\tilde U(2)$.

If our interpretation of the classical monopole solutions 
found in  the weak electric coupling regime as coherent states of 
the magnetic group $\tilde U(2)$ is correct, then there is presumably  
some fundamental
principle at work which forces the fundamental magnetic degrees of 
freedom to form coherent states in the weak electric coupling regime.
Since we do not have a formulation of Yang-Mills theory where  both 
electric and magnetic symmetry are manifest we are not able to identify
such a principle. In the absence of a deductive proof of our proposal we 
will therefore assess its  value by studying some of its consequences.

One immediate implication of the proposal is that there is 
 a natural mapping  from  the continuous magnetic orbits found
in the classification of classical monopoles  to finite-dimensional
 representation spaces of the magnetic group.
Mathematically this mapping is precisely the one one 
 would obtain according to the geometric quantisation prescription, 
but our physical interpretation
is different. Whereas geometric  quantisation would suggest that 
the magnetic orbits are classical phase spaces with quantised volume, 
 we interpret the continuous
parameters of the magnetic orbits as labels of the over-complete set
of coherent  states. By inverting the relation  \coherent\ we 
are thus in particular able to write down the basis of the carrier
space $V_{K,k}$ in terms of the coherent states, thus solving the long-standing
problem of isolating the fundamental magnetic degrees of freedom 
in the weak electric coupling regime:
\eqn\invert{
|K;k,m\rangle = \int \sin\beta d \beta d\alpha \,\, D^{k*}_{mk}(P) 
|K;k,\hat\bk\rangle.
}
In particular we deduce that an element $(e^{i\chi}, g)\in \tilde U(2)$ of the 
  magnetic group acts on $V_{K,k}$  according to
\eqn\magact{
D^{K,k}(e^{i\chi},g)|K;k,\hat \bk\rangle = 
e^{iK\chi} e^{i k \delta} |K;k,{\cal G}\hat \bk \rangle
}
where $\cal G$ is the $SO(3)$ matrix associated to the $SU(2)$  matrix $g$.
The angle $\delta$  depends on $g$ and $\hat \bk$ in a way which   
is quite complicated and not
very illuminating, see e.g. \threerus.  Note that if we start
with a reference  state $|K;k,(0,0,1)\rangle$ the  vector 
$\hat \bk= {\cal G} (0,0,1)^t$ sweeps out the magnetic 
orbits shown in Fig.~1 under the action 
of the magnetic group. Thus, although we generated  the coordinates 
$(\alpha,\beta)$   in \magvec\  originally by acting with the 
unbroken gauge group $U(2)$ in one point, we now find that we should 
interpret them as stemming from the action of the 
magnetic group $\tilde U(2)$. However, the magnetic group action
not only rotates the charge vector $\hat \bk$ but also generates the 
$K$- and $k$-dependent phase factors in \magact.

An interesting  check of our proposal is to see whether 
 it allows us to reproduce (and perhaps better understand)
the  fusion rule  for magnetic charges
discussed in \BS. According to that rule monopoles with 
 magnetic charges $(K_1,\bk_1)$
and $(K_2,\bk_2)$ can be    combined only if 
 the non-abelian 
components  are either parallel or anti-parallel.
Then the charges are added like vectors and the combined charge
automatically satisfies the Dirac quantisation condition if the individual
charges do. In the language of coherent states the natural way of ``fusing''
two states is to compute their tensor product. Here there is no reason to
impose any condition on the states to be multiplied.
The general formula is   complicated:
\eqn\cohprod{\eqalign{
|K_1;k_1 ,\hat \bk_1\rangle \otimes |K_2; k_2,\hat \bk_2\rangle& = \cr 
 \sum_{k=|k_1-k_2|}^{k_1+ k_2} \sum_{m=-k}^k &  \sum_{s=-k_2}^{k_2} 
D^{k_2}_{sk_2}(P_1^{-1}P_2) C^{k,s+k_1}_{k_1k_1,k_2 s} D^k_{m,s+k_1}(P_1)
|K_1+K_2;k,m\rangle,
}}
where $P_1$ and $P_2$ are $SU(2)$ matrices whose images under 
the Hopf projection \Hopf\ are   $\hat\bk_1$
and $\hat \bk_2$ respectively, and $C^{k,s+k_1}_{k_1k_1,k_2 s}$
is an $SU(2)$ Clebsch-Gordan coefficient.
After  expanding  $|K_1+K_2;k,m\rangle$
according to \invert\ the above formula shows that the product of two
coherent states of spins $k_1$ and $k_2$ is a superposition of coherent 
states with spins between $|k_1-k_2|$ and ${k_1+ k_2}$.
Rather remarkably one obtains in this way a closed multiplication rule
for ``vectors of quantised length''. 

If it is true that only coherent states of the magnetic group show
up in the weak electric coupling regime  we should be able to
 reproduce our classical fusion
rule by projecting out the coherent states in the tensor product.
More precisely we project  the RHS of \cohprod\ onto each  of the 
 carrier spaces $V_k$, $|k_1-k_2| \leq k \leq k_1+k_2$,
and check whether the projected state is coherent. 
In general this is  not  the case,  but it does happen
when $\hat \bk_1$ and $\hat \bk_2$ are parallel or anti-parallel.
In the former case, when $\hat \bk_1$ and $\hat \bk_2$ are
 equal to, say, $\hat \bk$,   the product state is coherent:
\eqn\oldfus{
|K_1;k_1,\hat \bk\rangle \otimes |K_2;k_2,\hat \bk \rangle =
 |K_1+K_2;k_1+ k_2, \hat \bk\rangle. 
}
In the  latter case, the  product state contains a coherent state
in the carrier space $V_k$ with the lowest possible spin $k=|k_1-k_2|$:
\eqn\oldfuss{
|K_1;k_1,\hat \bk\rangle \otimes |K_2;k_2, -\hat \bk \rangle =
 |K_1+K_2; k_1- k_2, \hat \bk\rangle  + \hbox{incoherent states},
}
where we assumed without loss of generality that $k_1\geq k_2$.
If $ |k_1- k_2|=0$ the product state should be interpreted 
as the unique singlet
state.
Thus we indeed recover the classical selection rule on monopole charges
which may be multiplied and also reproduce the classical results in 
the cases where the classical  multiplication is allowed.

\subsec{Non-abelian electric-magnetic duality}

The identification of  magnetic monopoles  with states
in UIR's of the   magnetic group  is a necessary condition for 
the validity of any  formulation of 
non-abelian  electric-magnetic duality. 
Having achieved this identification we can now go further and 
extend the Montonen-Olive electric-magnetic duality 
 conjecture to the case of non-abelian
unbroken gauge symmetry. Specifically in the theory we are considering
here we   propose that  the physics of purely electric 
particles  in  the  UIR $(N,j)$ of  $U(2)$  at coupling $e$ 
is the same  as that of purely magnetic particles in the UIR 
of $\tilde U(2)$ with labels $(K,k)=(-N,j)$  at coupling $4\pi /e$. 
The coupling constant $e$ is that of the electric formulation of the 
theory, so $4\pi /e$  should be thought of as the  coupling constant of a dual
or magnetic formulation of the theory. In particular  we therefore
 refer to the regime $e\ll 1$ as the weak electric coupling regime and 
to the regime $e\gg 1$ as the weak magnetic coupling regime.

One immediate
and interesting corollary of  the non-abelian electric-magnetic 
duality  conjecture  concerns the coherency requirement on electric
and magnetic degrees of freedom. Since no such requirement applies
to electric states in the weak electric coupling regime we deduce
that magnetic degrees of freedom  are not necessarily 
coherent in the weak magnetic coupling regime.
 Conversely it follows from our coherency postulate for magnetic states
in the weak electric coupling regime that in the weak magnetic 
(=strong electric) coupling regime only electric coherent states 
are allowed. In particular they would therefore have to obey
the selection rule for  tensor products, discussed in its magnetic
version above: only electric coherent states with parallel or 
anti-parallel charge directions may be combined in a tensor product.
 This last selection rule for electric states is 
reminiscent of a discussion by Corrigan
of   point-sources
for static Yang-Mills fields. In   \Corrigan\  it was  pointed out that 
only point sources with  non-abelian classical charges which lie on the 
same Weyl-orbit of the gauge group can be combined to produce
a static field. For the group $U(2)$ discussed here
two charges lie on the same  Weyl
orbit precisely if they have parallel or anti-parallel  non-abelian 
charge directions $\hat \bk$.

Non-abelian electric magnetic duality is also crucial in getting
a complete picture of the  possible charge sectors and their
fusion properties in Yang-Mills-Higgs theory with non-abelian
unbroken gauge group. Since this is one of the main concerns of this
paper we have devoted a separate section to it.

\newsec{Charge sectors and fusion rules }
 
In this paper we use the term ``fusion''
in a general sense to refer to the process of combining different 
 charge sectors of the theory.  In particle theory different
charge sectors are usually in one-to-one correspondence with 
UIR's of some group, and the 
answer to the fusion problem is then given by the Clebsch-Gordan series
of that group. However, it is well-known that in particular in two-dimensional
theories fusion properties are sometimes
  dictated by the representation ring
of other algebraic structures, such as quantum groups. Here we  will  encounter
yet a different situation: we are not able to sum up  
the fusion properties  in the representation ring  of a single algebraic
object. Instead the fusion properties depend on the coupling constant $e$,
with different groups classifying the sectors and organising the fusion
behaviour in the weak electric coupling regime   and the 
weak magnetic  coupling regime.
Only certain states which are present at all values of the coupling
 - the coherent states -
obey universal fusion rules.

We begin  our classification of the charge sectors with non-abelian
electric sectors, and initially consider the weak electric 
coupling regime.  Non-abelian electric sectors are  defined by the absence
of non-abelian magnetic charge  $k$.
States in this regime transform under the electric group $U(2)$  and can
be grouped into UIR's of that group. Such UIR's are labelled
by an integer $N$ and a positive half-integer  spin $j$ satisfying
$N+2j\in 2\bZ$. A basis of states
for this sector   is  furnished by the
tensor product  $|N;j,m\rangle  = |N\rangle \otimes |j,m\rangle$
of the $U(1)$  state $|N\rangle$  with  the customary basis states 
  $\{|j,m\rangle | m=-j,-j+1, ... ,j-1,j\}$
of the spin $j$ representation of $SU(2)$.
   If the abelian  magnetic charge $K$ is zero
the  sector is purely electric and 
 contains the familiar perturbative
massless and massive states. If the magnetic charge 
has the form  $(K\neq 0,k=0)$ (i.e. it
lies on one of the magnetic orbits on the vertical axis 
in Fig.~1.)  the integer $K$ is necessarily even
and  the excitations, studied in \BS,  are  then dyonic. Such dyonic sectors
are thus labelled  by the  triplet   $(K,N,j)$ and  for later use 
we introduce  basis states via
\eqn\elles{
|K;N;j, m\rangle =|K\rangle \otimes |N;j,m\rangle.
}
The fusion rules of  the non-abelian electric
 sectors are dictated by the representation
ring of the group $\tilde U(1)\times U(2)$: the abelian  magnetic charges 
simply add and the electric charges combine according to the 
familiar Clebsch-Gordan series of $U(2)$.

To extend our understanding of non-abelian electric charge sectors
to the strong electric coupling regime we use the corollary  of  non-abelian
electric-magnetic duality noted 
at the end of the previous section: in the strong electric coupling regime
only coherent states of the electric group $U(2)$ are physical. 
Considering without loss of generality the case of $K=0$,
the coherent electric states  are superpositions of the basis vectors 
$|N\rangle \otimes |j,m\rangle$ of the $U(2)$ UIR $(N,j)$ introduced above:
\eqn\elcoh{
|N;j,\hat \bk\rangle = |N\rangle \otimes \sum_{m=-j}^j D^j_{mj}(Q)|j,m\rangle.
}
The $SU(2)$ matrix  $Q$ is parametrised by  Euler 
angles $(\alpha,\beta,\gamma)$
and the direction of $\hat \bk$ is  given in terms of  $(\alpha,\beta)$ 
as in the formula \magvec\ for the magnetic charge direction. 
The use of the  same angular  coordinates
to parametrise  both magnetic and electric charge directions is no
accident  but a manifestation of the deep result that magnetic and 
electric symmetry are never simultaneously realised. Thus it is 
possible that, depending on the charge sector, 
 the same  coordinates  may have an 
electric  or a magnetic interpretation.
In \BS\  we saw for example   that 
in a fusion process  magnetic   parameters (generated by the action
of the magnetic group) acquire  an electric interpretation. 
We will return to such fusion processes after we have completed the
classification of charge sectors. Here we note that within the
non-abelian electric charge sector at strong electric coupling
the fusion rules are  the same as those in  the non-abelian magnetic
charge sectors at  strong magnetic coupling: only states with
parallel or anti-parallel charge directions $\hat \bk$ 
may be multiplied, and the result is
\eqn\elfus{\eqalign{
|&N_1;j_1,\hat \bk\rangle \otimes |N_2;j_2,\hat \bk\rangle  = 
|N_1+ N_2;j_1+j_2,\hat \bk\rangle    \cr
 |&N_1;j_1,\hat \bk\rangle \otimes |N_2;j_2,-
\hat \bk\rangle  = |N_1+N_2;j_1-j_2,\hat \bk\rangle + \hbox{incoherent states}.
}}

Next consider non-abelian magnetic charge sectors,
defined by the absence of non-abelian electric charge, i.e.  $j=0$.
 At weak electric coupling
 non-abelian magnetic  degrees of freedom
 necessarily form  coherent states \utwoco\ and obey the fusion rules
\oldfus\ and \oldfuss.  By duality   the requirement of 
coherency no longer applies
 in the weak magnetic coupling regime. 
Magnetic states are then arbitrary elements
of UIR's of the   magnetic  group $\tilde U(2)$. Such UIR's 
are labelled as before by
a pair $(K,k)$  of an integer $K$ and a positive half-integer $k$ satisfying
$K+2k\in 2\bZ$.  The abelian electric  charge  $N$ may be zero, in which case
the  sector is purely magnetic.  If the electric charge $N$
is even but non-zero  we have  dyonic sectors
 labelled  by the  triplet $(K,k,N)$, characterising  UIR's
of $\tilde U(2)\times  U(1)$. The fusion rules of states in these 
sectors  are given by  the Clebsch-Gordan coefficients
of  that group. Physical states in these sectors are tensor products
of the non-abelian magnetic states and abelian electric states.
At weak electric coupling they are magnetically coherent \utwoco:
\eqn\dycohe{
|K;k,\hat \bk;N \rangle = |K;k,\hat \bk\rangle \otimes |N\rangle.
}
At weak magnetic coupling a basis can be written in terms of the magnetic
states \invert:
\eqn\dycohh{
|K;k,m;N \rangle = |K;k,m\rangle \otimes |N\rangle,
}
and arbitrary  linear combinations of these basis states are allowed.

Finally we turn to non-abelian dyonic sectors,
described in great detail in \BS\ in the weak electric coupling 
regime. There  the non-abelian magnetic charge of a coherent
magnetic state obstructs the implementation of the electric group
$U(2)$. More precisely, on a monopole of charge $(K,\bk\neq 0)$
only that  subgroup of $U(2)$ can be implemented which leaves the 
magnetic charge invariant. In this case this is a maximal torus 
$T^2(\hat \bk)$
of $U(2)$ characterised by the  unit vector 
$\hat \bk$.
For a physical interpretation of the corresponding charges it is 
useful to separate the diagonal $U(1)$ subgroup of $U(2)$
and write the maximal torus as 
 $T^2 (\hat \bk) = (U(1)\times T^1(\hat \bk))/\bZ_2$,
where  $T^1(\hat \bk)$  is the torus subgroup  in  $SU(2)$ generated by  
$(\hat k_1 \tau_1 + \hat k _2 \tau_2 + \hat k_3 \tau_3)$ (the 
$\tau_i$ are the  Pauli matrices) and $\bZ_2$ is the group $\{1,-1\}$. 
 Dyonic quantum states
are characterised by giving the underlying monopole
state as in \mopost\ and then specifying the 
$(U(1)\times T^1(\hat \bk))/\bZ_2$ 
charges $(N,n)$, the former being an integer and the second a half-integer
such that $N+2n$ is even. Such dyonic states can thus be written as 
\eqn\dyst{
|K,k;N,n;\hat \bk\rangle.
}

By duality we  expect the following 
description of dyons to hold in the weak magnetic coupling regime. 
Now electric states are necessarily
coherent, and if they carry non-abelian electric charge it will
obstruct the implementation of the magnetic group $\tilde U(2)$.
More precisely on  an  electric  coherent state $|N;j,\hat \bk\rangle$
only that subgroup of $\tilde U(2)$  can be implemented which leaves 
$\hat \bk$  invariant. This is the  maximal torus $\tilde T^2(\hat \bk)$
which, in analogy with the electric torus $T^2(\hat \bk)$,
we rewrite  as  
 $\tilde T^2(\hat \bk) = (\tilde U(1)\times \tilde T^1(\hat \bk))/\bZ_2$.  
Denoting the eigenvalues  of 
the $\tilde U(1)$  and $\tilde T^1(\hat \bk)$ 
generators respectively  by $K$ and $k$,
a dyonic state in the weak magnetic coupling regime can be written
as $|K,k;N,j;\hat \bk\rangle$, which has precisely the same structure
as the weak electric coupling state \dyst. This similarity suggests that 
we could think  of dyonic states more symmetrically  as charged 
with respect to the subgroup
\eqn\dygroup{
\tilde T^2(\hat \bk) \times T^2(\hat \bk) \subset \tilde U(2) \times U(2).
}

Our discussion of the charge sectors can be summed up in the following
table:

\vskip 0.2 in
\def\bk{{\hbox{\bf k}}}
\vbox{\offinterlineskip
\hrule
\halign{&\vrule#&\strut\quad#\quad\hfil\cr
height5pt&\omit& &\omit& &\omit& &\omit& \cr
&\hfil Sector type &&\hfil Symmetry group  && 
States at $e\ll 1$ \hfil  && States at $e \gg 1$ & \cr
height5pt&\omit& &\omit& &\omit& &\omit& \cr
\noalign{\hrule}
height5pt&\omit& &\omit& &\omit& &\omit& \cr
&  electric && $\tilde U(1) \times U(2)$ \hfil && 
$|K;N;j,m\rangle$ && $|K;N;j,\hat \bk\rangle $ & \cr
&  dyonic && $\tilde T^2(\hat\bk)  \times T^2 (\hat \bk)$ \hfil && 
$|K,k;N,n;\hat \bk \rangle$ && $|K,k;N,n;\hat \bk\rangle $ & \cr
& magnetic && $\tilde U(2) \times U(1)$ \hfil && 
$|K;k,\hat \bk;N\rangle$ && $|K;k,m;N\rangle $ & \cr
height5pt&\omit& &\omit& &\omit& &\omit& \cr}
\hrule}

\bigskip
{\footskip14pt plus 1pt minus 1pt \footnotefont
\centerline{{\bf Table 1}: {\it Non-abelian charge sectors and their states}}
}
\bigskip

Remarkably the groups listed in table 1   are all subgroups 
of $ \tilde U(2) \times U(2)$ with the property that the two factors
centralise each other. This is the basis of the general statement 
made in the introduction that quantum states in Yang-Mills
theory with non-abelian unbroken gauge group $H$  can at most  be charged
with  respect to subgroups of $\tilde H $ and $  H$ which centralise each other
(when $H$ and $\tilde H$ are not
isomorphic - such as in the case of non-simply laced $H$ - the centralising
property should be defined in the adjoint representation).
It is satisfying that one can sum up the intricate interplay
between electric and magnetic charges in this neat,  general statement.
 In particular this result suggests that the following union of
commuting  pairs  in $\tilde U(2) \times U(2)$ 
\eqn\union{\eqalign{
U(2)_{\footnotefont {com}}=\left(\tilde U(2) \times  U(1) \right)
 \cup \left( \bigcup_{\hat \bk \in S^2}  
\tilde T^2(\hat \bk) \times T^2(\hat \bk) \right)  
\cup \left(\tilde U(1) \times U(2)\right) 
}}
plays a central role in understanding   the charge sectors of 
Yang-Mills-Higgs theory with unbroken gauge group $U(2)$.
We have used the notation $U(2)_{\footnotefont {com}}$ 
because this set can also
be defined as the  set of commuting pairs of elements in 
$\tilde U(2)\times U (2)$:
\eqn\hcom{
U(2)_{\footnotefont {com}}= \{(g,h)\in \tilde U(2)\times U(2) |gh=hg\}.
}
However, while this set has many interesting properties, it does
not appear to be endowed with a natural algebraic structure.
In particular it is not closed under multiplication.
The product of two elements in $U(2)_{\footnotefont {com}}$
   only belongs to $U(2)_{\footnotefont {com}}$
if the two elements both belong to one of the groups in the decomposition
\union. As we have seen, 
the Clebsch-Gordan series of those groups  tells us how
to combine two sectors of the same type (for example 
 two non-abelian electric sectors), but 
the structure of   $U(2)_{\footnotefont {com}}$ does not tell us 
how to combine 
a non-abelian electric state with a dyonic state, charged with respect
to $\tilde T^2(\hat \bk) \times T^2(\hat \bk)$ for  some $\hat \bk$. 
To answer  that question we need to combine results
from  our earlier paper \BS\  with non-abelian electric-magnetic 
duality.

The key result of  the  paper \BS\ is that
in the weak electric coupling regime  electric, 
magnetic, and dyonic states  can all be interpreted as carrying 
representations
of the semi-direct product group $U(2)\ltimes \bR^4$.
Here  the  group $\bR^4$ should be thought of as a ``magnetic
translation group'':  elements of  the  dual  $(\bR^*)^4$
are physically interpreted as magnetic charges.
As   vector spaces  $ \bR^4$  and the Lie algebra of $U(2)$
are isomorphic, and
$U(2)$  acts  on an element of $ \bR^4$ by conjugation of the 
associated element of  the Lie algebra
of $U(2) $.   We remind the reader that  representations 
  of semi-direct
products (like the Poincar\'e group or the Euclidean group)
are characterised by orbits and centraliser
representations. In the case of $U(2)\ltimes \bR^4$ the relevant
orbits are those of  $U(2)$ acting on  $(\bR^*)^4$.
If the orbit is trivial (one of the points on the central axis of Fig.~1)  
representations are
purely electric $U(2)$ representations. If the orbit is non-trivial
(one of the two-spheres in Fig.~1) representations are characterised by the 
two orbit labels $K$ and $k$  and an  UIR   of the 
$T^2 $ subgroup of $U(2)$ which leaves  a chosen point on 
the orbit invariant (centraliser representation). 
In \BS\ elements of  such representations were
called purely magnetic if the centraliser representation
is trivial  and dyonic otherwise. Moreover
we wrote down  bases for these representations which have precisely
the  magnetic \mopost, electric \elles\ and dyonic \dyst\ form described 
here, and which  can be realised as 
wavefunctions on monopole moduli spaces. The representation theory
of $ U(2)\ltimes \bR^4$ does not naturally select orbits with
the quantised ``height'' and ``radius'' found in the monopole spectrum
shown in Fig.~1, and in \BS\ this requirement had to be imposed by hand.
Similarly the condition 
that  the non-abelian magnetic charge directions $\hat \bk$ of two monopoles 
or dyons to be multiplied have to be equal or opposite is not natural in 
the context of $U(2)\ltimes \bR^4$ representations, and had to 
be imposed additionally. However,
with these restrictions the Clebsch-Gordan series 
of $U(2)\ltimes \bR^4$ resolves the difficult question of how
to combine states from different sectors. In particular it leads
to a formula for the tensor product of dyonic states with  equal
and opposite non-abelian magnetic charges as a superposition of dyonic states
 $|K;N;j,m\rangle$  in non-abelian electric
$\tilde U(1)\times U(2)$ representations $(K,N,j)$:
\eqn\tenabc{\eqalign{
 |K_1,k;N_1, n_1;&\hat \bk \rangle \otimes 
|K_2,k;N_2, n_2;-\hat \bk \rangle \cr
& = \sum_{j=|n_1-n_2|}^{\infty}\sum_{m=-j}^j \sqrt{2j+1}\,
D^j_{m(n_1-n_2)}(P) \,\,
|K_1+K_2;N_1+N_2;j,m\rangle,
}}
where $P$ is parametrised as in \euler\  and 
$(\alpha,\beta)$ are again the angles determining the direction
of $\hat \bk$ as in \magvec.

Here we have learnt to think of the magnetic states 
as coherent states of the magnetic group $\tilde U(2)$, which  immediately 
accounts for  the    quantisation of the magnetic charges 
$K$ and $k$. Furthermore the coherency requirement explains the
condition of parallel or anti-parallel charges in a tensor product.
However, while the introduction of the magnetic group and finally
 the discussion of  the mutually centralising subgroups \union\
of $\tilde U(2)\times U(2)$ leads to a  completely satisfactory
classification of the various charge sectors found in our theory,
 the semi-direct product  $U(2)\ltimes \bR^4$ is indispensable
 in discussing  fusion rules of different sectors. The
fusion rules derived from $U(2)\ltimes \bR^4$ are only valid 
in the weak electric coupling regime (which was the context of the 
discussion in \BS). Using electric-magnetic duality we deduce
that the fusion rules in the weak magnetic coupling regime are 
dictated by the dual semi-direct product 
$\tilde U(2)\ltimes \bR^4$, with elements of  $(\bR^*)^4$
now interpreted as electric charges. Dualising our description of  the 
UIR's of $U(2)\ltimes \bR^4$ the reader should  have no difficulty
in checking that the UIR's  of $\tilde U(2)\ltimes \bR^4$
contain the purely magnetic, coherent electric and dyonic 
states expected in the weak magnetic coupling regime.
Interpreting    the states in the weak magnetic coupling regime
as  representations of $\tilde U(2)\ltimes \bR^4$ has immediate
implications for the fusion rules governing these states.
States may again be multiplied  under the condition that 
the directions $\hat \bk$ characterising the 
non-abelian electric charge are equal or opposite, and then the 
outcome is determined by the Clebsch-Gordan coefficients of 
$\tilde U(2)\ltimes \bR^4$. The important point, anticipated 
in the opening paragraph of this section,  is that the
resulting fusion rules  are dual
but not equal to the fusion rules in the weak electric coupling 
regime.

In summary, the fusion rules within each type
of sector - non-abelian electric, dyonic and magnetic - 
 are governed by the Clebsch-Gordan series of 
  the  groups $\tilde U(1)\times U(2)$, $\tilde T^2(\hat \bk)
\times  T^2(\hat \bk)$ and $\tilde U(2) \times U(1)$,
 but to understand fusion
between different types of sectors  one needs to resort to the 
semi-direct products $U(2)\ltimes \bR^4$ and $\tilde U(2)\ltimes \bR^4$
at weak electric and weak magnetic coupling respectively.
However, even these groups do not tell one how to combine 
dyons whose  associated  charge directions $\hat \bk_1$ and 
$\hat \bk_2$ are  neither parallel nor anti-parallel.
Such a combination does not seem to lead to a well-defined  state.

\newsec{Fusion rules for coherent states and non-abelian S-duality}

The structure and fusion properties of charge sectors in 
Yang-Mills-Higgs theory with non-abelian unbroken gauge group
is clearly much more  intricate  than in the
abelian example outlined in the introduction.
While  the picture presented in the previous section enjoys
non-abelian electric-magnetic duality we have not yet said 
anything about the extension of electric-magnetic duality
to $S$-duality. In this section we will show how 
to do this for coherent states. 
These states are present in all coupling regimes. Moreover
it is straightforward  to  understand   the Witten effect on 
coherent states. Together with our formulation
of non-abelian electric-magnetic duality this leads to an
 implementation of $S$-duality on coherent states.
As in the abelian example discussed in the introduction,  
 there is a remarkable link between  $S$-duality and fusion
rules: the $S$-duality transformations  are  automorphisms of the 
fusion ring of coherent states. We  therefore begin our 
discussion with the fusion properties of coherent states.

Since coherent states are simply special states in the 
sectors  classified in the previous section, their fusion
rules can be derived from the result presented there.
Rather surprisingly, the  truncation of the representation
ring of $U(2)\times \bR^4$ to electric coherent states is the same 
 as the truncation of  the representation
ring of $\tilde U(2)\times \bR^4$ to magnetic coherent states.
It is in  that sense that  coherent states obey universal fusion rules.
Here we show that this truncated ring is the representation
ring  of a certain group 
associated to the  unoriented magnetic/electric
charge direction $\pm \hat \bk \in \bRP_2$. 
For simplicity we will assume in this section that $\hat \bk$
is the unit vector $(0,0,1)$, so that the tori $T^2(\hat \bk)$
and $\tilde T^2(\hat \bk)$ are the standard  maximal tori of 
$U(2)$ and $\tilde U(2)$, consisting of diagonal matrices;
we denote these standard tori simply by $T^2$ and $\tilde T^2$.
The  tori associated to  general directions $\hat \bk$
can be obtained from the standard tori by conjugation with
an appropriate $SU(2)$ matrix.

Consider for example purely electric coherent states 
and recall that  in the  electric strong coupling regime
only states  with equal or opposite charge directions may be multiplied.
In a given UIR $(N,j)$ of $U(2)$ the
coherent states with   charge direction  $\pm\hat \bk =\pm (0,0,1)$
 are precisely the states with highest and lowest $SU(2)$
weight. They can be obtained from  each other by acting with (the 
representative of) the element $\pmatrix{0 &1\cr 1&0} \in U(2)$.
Moreover they span an UIR of a subgroup of $U(2)$,
namely of the the semi-direct product group
\eqn\elcog{
S_2 \ltimes T^2,
}
where $S_2$ is the permutation group of two objects, realised 
canonically as a subgroup of $U(2)$ (with the non-trivial element
given above). Clearly we can analogously define a magnetic group
$S_2 \ltimes T^2$ whose UIR's contain magnetic coherent states
which may be multiplied. More interestingly we can combine 
these two groups to define the subgroup  
\eqn\chotwo{
Coh(2)= 
S_2\ltimes (\tilde T^2 \times T^2)
}
of $\tilde U (2) \times U(2)$. Here 
the permutation group  $S_2$ is realised  canonically
in  the diagonal $U(2)$ subgroup of $\tilde U(2) \times U(2)$. 
The  UIR's of \chotwo\  contain dyonic as well as electric and magnetic
coherent  states associated with the 
direction $(0,0,1)$. There are  one-dimensional and   two-dimensional
UIR's. On the  one-dimensional UIR's  the $S_2$ action is trivial and 
at most  the abelian central  subgroup $\tilde U(1)\times U(1)$ acts 
non-trivially.
These representations are thus  labelled by  
 integers $K$ and $N$ characterising UIR's of that  central subgroup.
The two-dimensional UIR's are more interesting for us. They carry 
additional  half-integer labels 
 $k$ and $n$ (characterising the transformation behaviour under the tori
$\tilde T^1$ and $T^1$)  such that $K+2k$ and $N+2n$ are even.
Moreover only the relative sign of $k$ and $n$ matters in  the labelling
of an UIR,  so that either  $k$ or $n$  can without loss of generality 
assumed to be positive. 
If we take $k$ to be positive we
may interpret it as the remnant of the non-abelian magnetic 
charge (and thus equal to radius of the magnetic orbits
found at    weak electric coupling).
If  further $n$ happens to be zero we are in the magnetic situation
 described above and have a natural identification of the 
UIR $(K,k,N,n=0)$ with the span of the two magnetic coherent states
associated with the direction $\hat \bk$.  If $n\neq 0$  there is 
a natural identification of the UIR $(K,k,N,n)$ with the span of 
of the  two   dyonic  states $|K,k;N,\pm n;\pm \hat \bk \rangle$ 
 discussed in the previous  section  \dyst. 
If, on the other hand,
 we take  $n$ to be positive we may 
 think of it as the remnant of the  the non-abelian  electric charge
(denoted $j$ in Sect.~4). Then, if $k$ happens to be zero
we are in the electric situation described above and identify  the 
UIR $(K,k=0,N,n)$ with the span of the electric coherent states associated 
with the direction $\hat \bk$. If $k\neq 0$
 we have a  natural identification of the UIR 
$(K,k,N,n)$ with the span of 
the two dyonic states $|K,\pm k;N,n;\pm \hat \bk\rangle$.

Having identified the group $Coh(2)$ as the algebraic object
whose representations classify coherent and dyonic  states associated with
a particular direction we show that  the fusion properties
of  these states  are encoded in  the representation ring of $Coh(2)$. 
For that purpose, and for a better understanding of the 
structure of $Coh(2)$,  a slightly more general perspective
is useful. We therefore  briefly consider the situation where
 the unbroken gauge group is $U(r)$. Then we define $Coh(r)$
analogously as a subgroup of $\tilde U(r)\times U(r)$: 
\eqn\chor{
Coh (r)=
S_r\ltimes (\tilde T^r\times T^r),
}
where $\tilde T^r$ and $T^r$ are the canonical maximal tori of $\tilde U(r)$
and $ U(r)$, and the permutation group   $S_r$ (the Weyl group of 
$SU(r)$) is realised canonically as a subgroup of the diagonal
$U(r)$ subgroup of $\tilde U(r)\times U(r)$.
The most natural tool for  discussing the Clebsch-Gordan
series  of this group  are characters, which are central  functions on 
the group. By definition this means that they only depend on the 
conjugacy class of an element and are therefore effectively functions
on 
\eqn\mrdef{
M_r= Coh(r) / {\footnotefont conjugation} =
 \hbox{ Sym}^r\left(U(1)\times U(1)\right).
}
Explicitly one can thus think of  $M_r$  as an unordered set  of  pairs  of 
angular coordinates $(\lambda_l,\omega_l)\in [0,2\pi)^2$ $l=1, .., r$.
As an aside we note that this space also happens to be the moduli space of 
flat $U(r)$ connections on a torus. This suggests not only 
how one should generalise the present discussion to other gauge
groups  but also establishes interesting links with algebraic 
geometry.

Returning to  our example $r=2$, it  is convenient to change 
to coordinates which explicitly refer to the diagonal rotations
$\tilde U(1)$ and $ U(1)$ and the Cartan subgroups $\tilde T^1$ and 
$ T^1$;
these are like ``centre of mass'' and ``relative'' coordinates: 
\eqn\newcoor{\eqalign{
\Lambda &={\lambda_1+ \lambda_2 \over 2}, \qquad \lambda =
{\lambda_1-\lambda_2} \cr
\Omega &= {\omega_1 +\omega_2 \over 2}, \qquad \omega ={\omega_1 -\omega_2}.
}} 
Then we can coordinatise  $M_2$ explicitly as 
\eqn\mrparam{
M_2 = \{(\Lambda,\Omega, \lambda,\omega)
 \in [0,2\pi)^2\times [-2\pi,2\pi)^2\}/\sim
}
where the equivalence relation $\sim$ identifies
$(\Lambda,\Omega,\lambda, \omega) $ with 
$(\Lambda,\Omega,-\lambda, -\omega)$
and $(\Lambda,\Omega,\lambda,\omega)$ with $ (\Lambda +\pi,
\Omega + \pi, \lambda+2 \pi,\omega+2\pi)$.

 The character of  the 
$Coh(2)$-representation $(K,k,N,n)$  is then the  function 
\eqn\charr{
\chi_{K,k,N,n}(\Lambda,\Omega, \lambda,\omega) =
e^{i(K\Lambda +N\Omega)} \cos(k\lambda +n\omega).
}
It follows from
the general theory of characters (and can easily be checked explicitly)
that the set of $Coh(2)$  characters form an orthonormal basis of $L^2(M_2)$.
Then the  Clebsch-Gordan series of $Coh(2)$ can be read of from 
pointwise multiplication of the characters and subsequent expansion
in the  basis
$\chi_{K,k,N,n}$. This  yields  the following fusion rules 
for the 
sectors $(K,k,N,n)$:
\eqn\fusion{\eqalign{
(K_1,k_1,N_1,n_1) \otimes (K_2,k_2,N_2,n_2)
= (K_1&+K_2,k_1+k_1, N_1+N_2, n_1+n_2)\cr  & \oplus
 (K_1+K_2,k_1-k_1,N_1+N_2, n_1-n_2).
}}
These are  the promised  ``universal''  fusion rules for coherent 
and dyonic states
associated with a given  unoriented  vector  $\pm \hat \bk$.
In particular they agree with the fusion rules one obtains 
from tensoring coherent  or dyonic states according to    
the representation theory of 
 $U(2)\ltimes \bR^4$  or $\tilde U(2)\ltimes   \bR^4$
(with the (anti-)parallelity  condition  on  the charge direction  $\hat \bk$)
and  subsequently truncating  to coherent and  dyonic  states.

The challenge of formulating  non-abelian $S$-duality consists of combining
our formulation of electric-magnetic duality with
an implementation of the   
Witten effect such that the two generate  an $SL(2,\bZ)$ action.  
In this wording of the task 
it is a priori not even clear whether we should aim for  an
action on charge sectors or on individual quantum states
(in the abelian situation where all UIR's are one-dimensional
these two possibilities coincide).  Here we are going to propose
an $SL(2,\bZ)$-action which maps  the  coherent states
in one charge sector onto the coherent states of another.
The central role of coherent states stems from the fact that 
they are present  in all coupling regimes. 
A practical advantage is that we have no difficulty implementing
the  Witten effect  on coherent states. For suppose we have
a purely magnetic coherent state $|K;k,\hat \bk\rangle$. In the weak  electric
 coupling regime we identify it with a classical monopole
with charge $(K, \bk)$. A simple extension of the 
original calculation performed by Witten \Wittenef\ shows that 
a shift in the $\theta$-angle by $2\pi$  transforms this state into
a dyon as follows:
\eqn\witef{
|K;k,\hat \bk\rangle \rightarrow |K,k;K,k;\hat \bk\rangle. 
}

The key to the implementation of full $S$-duality  on coherent
states is a 
 natural $SL(2,\bZ)$ action on 
 $M_2$. In order to indicate the generality of the 
construction we write down this action  for $M_r$.
An element of $SL(2,\bZ)$, written  as in \sltwoz, acts on each pair
$(\lambda_l,\omega_l)$ in the parametrisation \mrdef\ of $M_r$  according to
\eqn\nonabs{
(\lambda_l,\omega_l) \rightarrow
 (d \lambda_l -b \omega_l,-c \lambda_l +a \omega_l).
}
Since the action is the same for each pair it clearly
commutes with the action of the permutation group $S_r$ on the set of pairs
$\{(\lambda_\rho,\omega_\rho)\}_{\rho =1, ..., r}$ 
and is thus well-defined on $M_r$.
The action of the modular group on the manifold $M_r$ induces
an action on functions on $M_r$ and in particular therefore on characters
of the group $Coh(r)$. We explicitly describe this again in the case
$r=2$.

 In that case
we find   in particular  the following action of the  generator $S$ of 
 electric-magnetic duality
\eqn\sssdu{
(K,k,N,n) \rightarrow (-N,-n,K,k),
}
which should be combined with the inversion of the complex coupling 
constant $\tau \rightarrow -1/\tau$. Recall that only the 
 relative sign of $k$ and $n$ matters in the labelling of $Coh(2)$ 
representations; the action of $S$ changes this relative sign.
The generator $T$ implements the Witten effect in the required  fashion
\eqn\tttdu{
(K,k,N,n) \rightarrow (K,k,N+K,n+k).
}
While these formulae are very similar to the formulae given for abelian
$S$-duality in Sect.~1 we emphasise that they refer to (in general) 
two-dimensional UIR's of $Coh(2)$. Applied to non-abelian  magnetic states
for example \sssdu\ maps a doublet of magnetic coherent states onto
a doublet of electric coherent states and \tttdu\ maps a doublet
of magnetic coherent states onto a doublet of dyonic states. 
In the more general case of  $U(r)$ as unbroken gauge group our
formalism would yield a map between  higher-dimensional 
(at most $r$-dimensional)
sets of magnetic, electric and dyonic states. These sets are in one-to-one
correspondence with  orbits  of  the Weyl group $S_r$ of $SU(r)$.
Weyl orbits have played a role in earlier discussions of duality \GNO\
and here we find them back as sets of coherent and/or dyonic states
which may be multiplied consistently at all values of the coupling $e$.  

Finally we note that, as in the abelian case,  
$S$-duality transformations  are 
automorphisms of the  fusion rules. 
This follows automatically from our encoding of
the fusion rules in the pointwise multiplication of characters which 
commutes with the $SL(2,\bZ)$-action on the arguments of the characters.

\newsec{Conclusion}

The realisation of symmetry and the implementation
of $S$-duality in Yang-Mills theory with non-abelian residual symmetry
is richer and   more intricate than in the abelian situation.
In this concluding section we  highlight three important
general points  of our discussion.

The first point concerns the charge sectors. 
While both electric and magnetic non-abelian symmetry can be 
found in the theory,  they are never fully  realised at the same time.
The allowed charge sectors are classified by UIR's of  pairs of
commuting subgroups of the electric and magnetic symmetry groups,
as displayed explicitly in \union. 

The second point concerns the fusion rules. 
Within a given type of sector
these are given by the representation ring of a group.
However fusing different types of sectors is complicated and depends
on the coupling regime. In our case, fusion rules were encoded in
the representation ring of the 
 semi-direct product group $U(2)\ltimes \bR^4$ at weak electric
coupling but in the representation ring of the 
 dual $\tilde U(2)\ltimes  \bR^4$
at weak magnetic coupling.

The third point concerns duality. Our classification
of sectors and their fusion rules  enjoy  manifest  non-abelian
electric-magnetic duality, but to implement $S$-duality 
we restricted attention to coherent states. 
These are special in that they appear in all 
coupling regimes and have universal fusion properties
in the sense that their fusion ring is a sub-ring of
the representation  ring of both $U(2)\ltimes \bR^4$ and
   $\tilde U(2)\ltimes  \bR^4$. 
On  coherent states  $S$-duality can be implemented  as an $SL(2, \bZ)$
action on charge sectors  which leaves the  fusion rules invariant.
Conversely one could define $S$-duality as the automorphism
group of the  fusion ring of coherent states.

There are a number of further questions which arise from our discussion.
At the technical level one would like to generalise to other gauge
groups, particularly  ones which are not simply laced. At a deeper 
conceptual level one would like to find a reason for the ``freezing'' 
of magnetic (electric)  degrees of freedom into coherent states at
electric  (magnetic) weak coupling. One would also like to say 
more  about  the intermediate range $e\approx 1$,
where  fusion properties are   captured by neither the electric 
$U(2)\ltimes \bR^4$  nor  the magnetic $\tilde U(2)\ltimes  \bR^4$.
While we are confident to be able to report on the technical issue
of general gauge groups in the near future,  the  conceptual
questions  pose a deeper challenge.

\ack{We are grateful to Erwyn van der Meer for providing figure 1
and to  Robbert Dijkgraaf, Klaas Landsman,
Nathalie Muller, Erik and 
Herman Verlinde for discussions.
BJS also thanks Sharad Agnihotri,
Fred Goldhaber, Kimyeong Lee, Martin Rocek, Cumrun Vafa
and Erick Weinberg   for useful comments
and  acknowledges financial support through a  Pionier Fund of 
the Nederlandse Organisatie voor Wetenschappelijk Onderzoek (NWO). 
}

\listrefs
\bye